\documentclass[sigconf, screen]{acmart}
\AtBeginDocument{%
  \providecommand\BibTeX{{%
    \normalfont B\kern-0.5em{\scshape i\kern-0.25em b}\kern-0.8em\TeX}}}

\setcopyright{acmcopyright}
\copyrightyear{2022}
\acmYear{2022}
\acmDOI{XXXXXXX.XXXXXXX}

\usepackage{graphicx}
\usepackage{amsmath}
\usepackage{amsthm}
\usepackage{booktabs}
\usepackage{multirow}
\usepackage{adjustbox}
\usepackage{enumitem}
\usepackage{siunitx}
\usepackage{amsfonts}
\usepackage{subfigure}
\newcommand\Text[0]{{\it text}}
\newcommand\hist[0]{{\it hist}}
\newcommand\N[0]{\mathcal{N}}
\newcommand\SN[0]{\mathcal{G}}
\newcommand\M[0]{\mathcal{M}}
\newcommand\E[0]{\mathcal{E}}
\newcommand\V[0]{\mathcal{V}}
\newcommand\VA[0]{\mathcal{L}}
\newcommand\ppm[0]{\scriptstyle\pm}
\newcommand\paratitle[1]{\vspace{1mm}\noindent\textbf{#1}}

\DeclareMathOperator*{\argmax}{argmax}

\setcopyright{none}

\begin{document}

\title{``Double vaccinated, 5G boosted!":
Learning Attitudes towards COVID-19 Vaccination from Social Media}
\author{Ninghan Chen}
\affiliation{%
 \institution{University of Luxembourg}
 \streetaddress{L-4364}
 \city{Esch-sur-Alzette}
 \country{Luxembourg}}
\email{ninghan.chen@uni.lu}

\author{Xihui Chen}
\affiliation{%
 \institution{University of Luxembourg}
 \streetaddress{L-4364}
 \city{Esch-sur-Alzette}
 \country{Luxembourg}}
\email{xihui.chen@uni.lu}

\author{Zhiqiang Zhong}
\affiliation{%
 \institution{University of Luxembourg}
 \streetaddress{L-4364}
 \city{Esch-sur-Alzette}
 \country{Luxembourg}}
\email{zhiqiang.zhong@uni.lu}

\author{Jun Pang}
\affiliation{%
 \institution{University of Luxembourg}
 \streetaddress{L-4364}
 \city{Esch-sur-Alzette}
 \country{Luxembourg}}
\email{jun.pang@uni.lu}

\begin{abstract}
To address the vaccine hesitancy which impairs the efforts of the COVID-19 
vaccination campaign, it is imperative to understand public vaccination 
attitudes and timely grasp their changes. 
In spite of reliability and trustworthiness, conventional attitude collection 
based on surveys is time-consuming and expensive, and cannot follow the 
fast evolution of vaccination attitudes.
We leverage the textual posts on social media to extract 
and track users' vaccination stances in near real time by proposing a deep learning framework.
To address the impact of linguistic features
such as sarcasm and irony commonly used in vaccine-related 
discourses, we integrate into the framework the recent posts of a user’s social network neighbours 
to help detect the user’s genuine attitude. 
Based on our annotated dataset from Twitter, the models instantiated from our framework
can increase the performance of attitude extraction by up to 23\% compared 
to state-of-the-art text-only models. 
Using this framework, we successfully validate the feasibility 
of using social media to track the evolution of 
vaccination attitudes in real life. 
We further show one practical use of our framework by validating the possibility to 
forecast a user's vaccine hesitancy changes with information perceived from social media. 
\end{abstract}



\maketitle

\section{Introduction}
\label{sec:introduction}

Vaccination is now unanimously accepted as an effective approach to 
combat the ongoing global COVID-19 pandemic,
caused by the contagious coronavirus SARS-CoV-2~\cite{KM21}. 
Despite the decreased efficacy against the infection of the
recent variants, 
a high-level uptake of the currently available vaccines is still believed as key to 
restrain the numbers of severe diseases, 
deaths, and particularly hospitalisation, which is crucial for medical systems 
to remain operating as normal~\cite{AS22}.  
Regrettably, similar to the vaccines of other infectious diseases, 
not everyone is willing to be vaccinated~\cite{DLG13}.
The impact of \emph{vaccine hesitancy} has been widely recognised and extensively
studied in a number of countries 
for various groups of people, e.g., healthcare workers 
and immigrants. 
Many related factors and their influences are evaluated and compared,  
e.g., education, income and gender~\cite{CPA21}. 
These scientific findings have provided policymakers with valuable references to 
design strategies to reduce vaccine hesitancy and 
fix the stagnant uptake ratios.

The success of these studies relies on 
the collection and accurate understanding of the public's attitudes towards vaccination. 
Social surveys with well-defined questions, due to   
their reliability and trustworthiness, 
have been adopted as the dominant source of public opinions in the literature. 
However, as conducting surveys is expensive and time-consuming, they tend to fall behind the fast development of the COVID-19 pandemic~\cite{MMnature20}, 
and thus fail in capturing the evolution of vaccine hesitancy. 
Continuous tracking of public vaccination attitudes 
can help healthcare bodies to identify the significant fluctuations 
to make a timely intervention or fast capture the public's response to 
implemented policies. 
Moreover, it allows for analysing  dynamic factors 
such as occasional social protests and misinformation propagation, 
in addition to the static ones like demographic profiles which rarely change in short periods. 

In recent years, social media has attracted the attention of data analysts as an  
auxiliary data source to complement public health surveillance (PHS)
in spite of its inherent bias, e.g., regarding population sampling~\cite{MMnature20,ARZ20}.   
Due to social distancing and fear of the unknown,  
people spend more time than ever on social media.
As social media posts have proved to encode posters' subjective opinions~\cite{johnson2020online,GCH19},
in this paper, we aim to leverage the enormous daily 
posts during the COVID-19 pandemic to extract users' vaccination attitudes and track their temporal changes. 

We take advantage of the recent advances of deep learning in natural language 
processing (NLP), and propose a framework that can accurately classify a textual post 
according to the vaccination stance expressed by its originator. 
Our framework effectively addresses the challenge impairing existing NLP models' 
performance, i.e., the linguistic features such as sarcasm and irony, which are 
quite common in vaccination-related discourse. 
Consider the following example: ``\emph{I wouldn't do it for their vaccine, 
I'm waiting for the 6G}''. The user expresses his/her support for vaccination
by making fun of the conspiracy theory that 
chips are implanted with vaccine injection.  
After experimenting with the state-of-the-art text feature-based 
classification methods, we only get an accuracy of $0.65$, which is apparently not 
reliable enough for trustworthy analysis. 
Recent studies revealed that a user's vaccination attitudes correlate with 
those of their neighbours in social networks, e.g., friends and friends of friends.  
For example, online social network users with negative attitudes often have 
social relations with users of positive attitudes~\cite{johnson2020online,morales2021no}.
Inspired by such studies, we integrate the recent posts of a user's 
social network neighbours in our framework to help detect the user's genuine attitude and 
reduce the impact of sarcasm. 

To train and test models instantiated from our framework, we collect 9,135,393 
tweets from Twitter generated by 69,936 users, and create the first 
annotated dataset of 18,246 tweets manually labelled with affective vaccination 
stances (e.g., positive and negative). 
In addition to the experimental evaluation, we draw the temporal evolution of 
vaccination attitudes extracted from our collected tweets. 
We cross-validate with published social studies and manually 
analyse the popular social events occurring around significant changes of 
vaccine hesitancy levels. 
All the validation results successfully illustrate the effectiveness of our framework, 
as well as the power of social media as a data source to grasp 
public vaccine hesitancy in practice in near real time. 

Newsagents, governments, healthcare professionals and even 
anti-vaccine activists use social media to spread news, knowledge and 
suggestions to persuade or dissuade people from getting vaccinated~\cite{BBB20}.
To showcase the practical use of our post-based attitude learning framework,  
to the best of our knowledge, we are the first to 
demonstrate that the information that users perceive from social media
can be used as predictors of their vaccine hesitancy changes. 

\paratitle{Our contributions.} Our contributions are as follows:
\begin{itemize}
    \item We propose a framework to extract vaccination stances from textual social media 
    posts.  Our framework integrates recent posts of a user's social network neighbours to 
    help reduce the interference of linguistic features, e.g., sarcasm and irony. 
    \item We design models instantiating our framework. 
    Based on our annotated dataset from Twitter, the best model 
    can increase the performance of attitude extraction by up to 23\% compared 
    to state-of-the-art text-only models. 
    \item Using the model with the best performance, we track the evolution of 
    vaccination attitudes. The utility of the extracted vaccination attitudes
    is further validated by the consistency with published statistics and 
    explainable significant fluctuations of vaccine hesitancy in terms of 
    social events such as wide propagation of misinformation and negative news.
    \item We show a practical use of our framework by validating the possibility to 
    predict a user's vaccination hesitancy changes with the information he/she perceives 
    from social media. 
\end{itemize}
Through this paper, we (re-)establish the power of social media as 
a complementary data source in public health surveillance in spite of its inherent biases. 
Specifically, when exploited properly, it can provide healthcare bodies with useful 
information to guide or support  their decision-making processes. 

%

\section{Related Work}
\label{sec:related-work}
Vaccine hesitancy is believed to be a major cause of stagnant vaccine coverage 
and contributor to vaccine program failure~\cite{DLG13}. 
In spite of the lack of a unified definition, one widely accepted representation 
of vaccine hesitancy is a continuum, ranging from complete rejection of 
vaccines to varying degrees of scepticism~\cite{wilson2014opportunities}.  
In this section, we concentrate on the vaccine hesitancy studies after the onset of 
the COVID-19 pandemic. 
%
%
A considerable amount of literature has been published investigating 
the state of vaccine hesitancy and the influential factors
in different regions~\cite{khubchandani2021covid,sherman2021covid} 
for specific groups of people such as healthcare employees~\cite{biswas2021nature,grech2020vaccine}, 
immigrants~\cite{alabdulla2021covid}  
and college students~\cite{barello2020vaccine}. 
Although surveys are still the most adopted method to collect sampled populations' attitudes or 
stances toward vaccination~\cite{AMM2021}, 
some recent works leverage social media as a new dimension~\cite{johnson2020online,GCH19}.
Compared to self-reporting questionnaires, 
social media data are cost-effective to access, 
and more importantly, allow analysis over large populations which was not previously 
feasible~\cite{MMnature20,ARZ20,johnson2020online}. 

The methods extracting vaccination attitudes from social media fall into two categories:
community-based and post-based. 
\citet{cossard2020falling} found pro- and anti-vaccine users naturally cluster 
into communities and calculated community partitions of various communication graphs 
to infer users' vaccination stances.
\citet{johnson2020online} made use of the topics of fan pages (similar to 
discussion groups) on Facebook to approximate users' attitudes, and analysed the 
communities formed by 100 million users across the world in terms of their vaccination attitudes. 
Post-based methods benefit from the various types of information encoded in social media
posts such as texts, labels and pictures. 
\citet{GCH19} relied on the hashtags in tweets to approximate the 
vaccination attitudes in tweets. Sentiment analysis, as part of natural language processing (NLP),  
aims to derive the subjective opinions expressed in texts.
The introduction of deep learning leads to more powerful models that can process posts 
at the sentence or paragraph levels such as word2vec~\cite{word2vec13} and BERT~\cite{DCLT19}. 
The sentiments extracted from texts have been used as references to study vaccine hesitancy~\cite{GADN21}.
For instance, \citet{GADN21} detected the opinions of media towards vaccines 
in Africa through Twitter and Google news.

\paratitle{Discussion.} The community-based methods cannot capture the fast development
of public vaccination attitudes due to the relatively stable connections between users. 
Moreover, community memberships are effective for analysis on the level of populations but 
fail to accurately derive individual users' attitudes.
The post-based methods in previous studies are not specifically 
designed and trained for COVID-19 vaccines. As a result, they cannot 
capture the special linguistic characteristics of the online discourses during the COVID-19 pandemic.
This is partially because of the lack of social media posts which are related to COVID-19 vaccines 
and annotated with vaccination stances. In this paper, we propose a framework that 
can not only benefit from state-of-the-art post-based methods, but also deal with the
interference of linguistic features such as sarcasm and irony in discourses related to 
COVID-19 vaccination. We also create the first annotated dataset of tweets which can facilitate 
developing future models on subjective opinion extraction. 

\section{Extracting Vaccination Attitudes from Social Media Posts}
\label{sec:attitude detection}


\subsection{Problem definition} 
Extracting the vaccination stance of a social media post 
can be technically formulated as a natural language processing 
(NLP) task~\cite{LPLXYSYH20}, i.e.,  
classifying texts  according to given class labels.
In this paper, we focus on the affective stance towards COVID-19
vaccination. Thus the set of labels is $\VA=\{ {\it NE}, {\it PO}, {\it NG}\}$ where 
${\it NE}$, ${\it PO}$ and ${\it NG}$ correspond to neutral, positive and negative, 
respectively.
The basic idea of text classification in NLP is first to calculate a 
representation of the given text which summarises 
its linguistic features and then output the most likely class label. 
Classification methods differ from each other in terms of the formats of text representation and 
the mapping from representations to class labels. 
Text classification confronts the challenge that the attitude or emotion expressed by 
the same words varies according to the context. 
For instance, the figurative usage of symptom words can fool the keyword-based 
classification methods and significantly deteriorate the precision 
of health mention classification~\cite{BJLPX20}. 
Sentiments of the short texts with symptom words are thus introduced 
as auxiliary information to deal with this figurative interference. 

As discussed previously, we notice that during the COVID-19 pandemic, 
Twitter users tend to use sarcastic or ironic expressions to 
describe their disagreements with those with different vaccination stances. 
Inspired by previous works such as~\cite{BJLPX20}, 
given a post, we aim to benefit from the originator's
social relations as well as their past posts to help reduce or eliminate 
the interference of sarcasm and irony.

We use $\SN=(\V,\E)$ to represent the social graph recording the 
social relations between users 
where $\V$ is the set of nodes and $\E\subset \V\times\V$ is the set of 
edges between nodes. A node $v\in V$ corresponds to a social media user
and an edge $(v,v')$ indicates the existence of a relationship between two users $v$ 
and $v'$. Note that we ignore the direction of relationships in this paper
to take into account all the neighbours of a user, e.g., both 
followers and followees on Twitter.
Thus, 
$(v,v')\in \E$ implies $(v',v)\in \E$.
We abuse the terms \emph{user} and \emph{node} in the following discussion 
when it is clear from the context. 
Let $\N_i^k$ be the set of neighbours of node $v_i$ within $k$ hops,
i.e., $\{v\mid d_\SN(v, v_i)\le k\}$ where $d_\SN(v,v_i)$ is the shortest distance 
between $v$ and $v_i$ in the graph $\SN$. Note that node $v_i$ is also 
in $\N_i^k$ as $d_\SN(v_i, v_i)=0$.
We use $x_i^t$ to denote the textual message posted by user $v_i$ at time $t$.
We use $\M^{< t}_i$ to denote the list of posts originated by user $v_i$
before time $t$  chronologically ordered by their post time,
and $\M^{< t}_{\N^k_i}$ to represent the set of post lists of the 
neighbours of user $v_i$ within $k$ hops. 

Our COVID-19 vaccination attitude classification problem can be defined as calculating 
the probability distribution of all labels in $\VA$. The final vaccination 
stance of $x_i^T$, i.e., $f(x_i^t)$, is determined by the label with the largest
probability. Formally, we have  
\[
f(x_i^t)=\argmax_{{\it stance}\in\VA} \Pr({\it stance}\mid x_i^t, \SN,  \M^{< t}_{\N^k_i}).
\]

\begin{figure}[!t]
\includegraphics[width=0.85\linewidth]{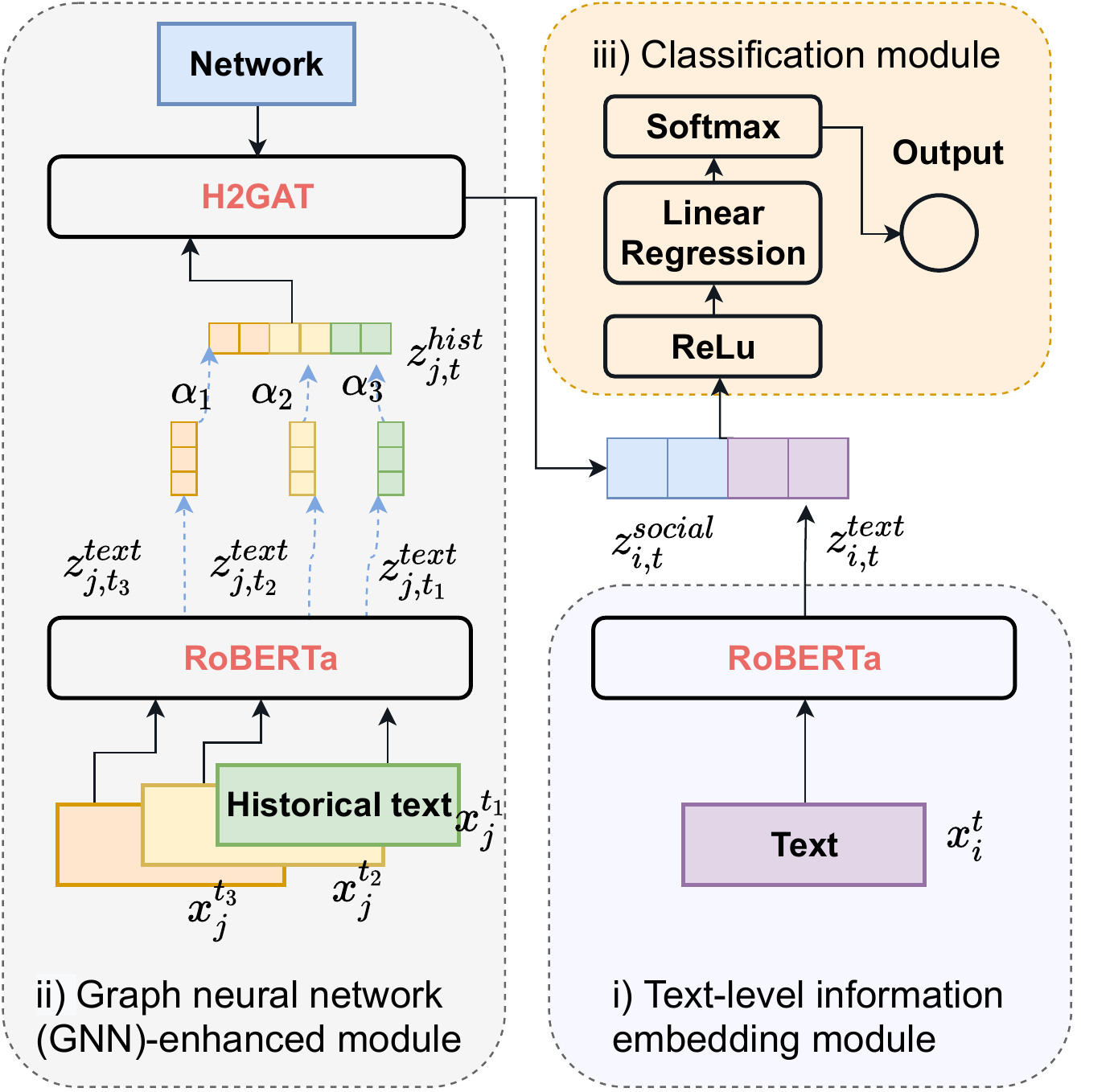}
\caption{An illustration of our attitude classification framework and model.}
\label{fig:model_architecture}
\vspace{-4mm}
\end{figure}

\subsection{A vaccination attitude learning framework} 
To solve the classification problem formulated previously, 
we propose a framework which takes advantage of the 
recent success of adopting deep learning in NLP and graph analysis such as text embedding 
and graph neural networks (GNNs).

Figure~\ref{fig:model_architecture} depicts an overview of our framework by labelling  
its three main components in different background colours: i) a text-level information 
embedding module, ii) a GNN-enhanced module, 
and iii) a classification module.
The first module is used to learn the linguistic representation of the targeted post 
while the second module summarises the linguistic features of the recent messages 
posted by the user's neighbours. 
We concatenate the outputs of these two modules as the input of the classification module.
GNN~\cite{KipfW17} has shown its advantage in transforming graph information, 
including structures and attributes of nodes and edges in both academia and industry. 
Intuitively, it employs a message passing scheme to integrate the information 
of a node's neighbourhood as the representation of the node. 
The calculated embedding is then used for 
various downstream applications such as node classification and link prediction. 
Variants of GNN differ from each other in terms of the 
implementation of their message passing schemes. 

Our framework can be instantiated by assembling various methods that can achieve 
the corresponding tasks of the modules. 
In the following, we present how we implement every module of the framework
and justify our choices.

\subsection{Our model}
\label{subsec:learn_from_text}

\noindent\textbf{Text-level information embedding.}
In order to mine meaningful information from a limited number of annotated posts, 
recent solutions leverage pre-trained NLP transformers to calculate the embedding for 
short texts~\cite{LFXYWJLX20}. 
NLP transformers 
have been empirically evaluated in~\cite{ZLZZ21} where RoBERTa~\cite{LOGDJCLLZS19} is shown to 
outperform the competing models.
Due to its high effectiveness, we adopt RoBERTa to learn text representations in our model. 
The model takes a textual post, e.g., $x_i^t$, as input, and 
outputs a sentiment-oriented representation vector 
$z^{\Text}_{i,t}\in\mathbb{R}^{d}$ where $d$ is the pre-defined dimension of the vector.
RoBERTa will be fine-tuned with the posts in the training set. 

\paratitle{GNN-enhanced module.}
Given the target post $x_i^t$, we utilise this module to capture the linguistic features 
of the recent messages posted by $v_i$'s friends within $k$ hops before $t$. 
The definition of ``recent'' is flexible and depends on application scenarios. 
In this paper, we select the last $\lambda$ tweets before $t$ as a user's recent tweets. 
Its output will be subsequently used as complementary contextual information to 
further ameliorate classification performance. 
Therefore, the input of this module consists of the social network  $\SN$ 
and the historical messages of  user $v_i$'s $k$-hop neighbours, i.e., $\M_{\N^k_i}^{<t}$. 
Note that the post originator's recent messages are also considered 
as $v_i$ is included in $\N_i^k$.
The output will be a text-level embedding vector that can be intuitively interpreted 
as a summary of useful features of friends' recent discourse.  

We take two successive steps to calculate the output text-level representation vector $z_{i,t}$. 
We first integrate the recent posts of each user in $v_i$'s $k$-hop neighbourhood 
as a summary of  his/her vaccination discourse.
In the second step, we propose a new GNN-based model named by H2GAT 
to aggregate the discourse of all $v_i$'s neighbours into the  social text encoding vector. 

\paratitle{STEP 1: Text-level encoding.}
For each user $v_j\in\N_i^k$, we use his/her last $\lambda$ posts in $\M_j^{<t}$,
denoted by the list $(x^{t_\lambda}_j, x^{t_{\lambda-1}}_j,\ldots, x^{t_1}_j)$ 
where $t_m$ ($1\le m\le \lambda$) is the time stamp of $v_j$'s last $m$-th post. 
We apply the pre-trained RoBERTa model and then obtain the corresponding list 
of text-level representations, i.e., $(z^{\Text}_{j, t_\lambda}, z^{\Text}_{j,t_{\lambda-1}}, \ldots, z^{\Text}_{j, t_1})$. 
There are many ways to integrate $v_j$'s past text-level representations while distinguishing 
their various temporal importance, e.g., Hawkes~\cite{Hawkes71} and GRU~\cite{CMBB14}. 
In our implementation, we adopt the dynamic-aware \textit{position encoding} by assigning 
a fixed importance factor to each past post according to its position in the list. 
This method is simple but more effective than other competing ones in our  experiments
(as shown in Section~\ref{sec:experiments}). 
Formally, the integrated text-level representations of user $v_j$ is calculated as follows:
$$z^{\hist}_{j,t} = \sum_{m\le \lambda} \alpha_m \cdot z^{\Text}_{j,t_m}$$
where $\alpha_m$ ($1\le m\le \lambda$) is trainable and describes the 
positional relation between the past posts. 
Note that $k$ and  $\lambda$ are predefined hyper-parameters that should be tuned manually. 

\smallskip\noindent
\textbf{STEP 2: H2GAT.}
It is pointed out that the \textit{heterophily} phenomenon widely 
exists among online social network users~\cite{pandit2007netprobe}. 
This phenomenon also exists in vaccination-related discourses as 
the attitudes and linguistic features can be significantly different between users. 
Considering the \emph{heterophily} of vaccination discussion in a user's 
neighbourhood, we adopt and extend a recent GNN-based method called H2GCN~\cite{ZYZHAK20}.
The same as other GNNs, it also has multiple layers, the $\ell$-th of which can be 
formulated as follows:
\begin{displaymath}
\begin{array}{ll}
    H^{\ell}_{i} = {\sf Combine}(\{{\sf Aggregate}\{H^{\ell-1}_{j}: 
    j \in \widehat{\N}^{k'}_{i}\}:   k' \in  \{1, \dots, k\} \})
\end{array}
\end{displaymath}
where $\widehat{\N}^{k'}_i$ represents node $v_i$'s $k'$-order neighbours, i.e., 
nodes that have an exact distance of $k'$ from $v_i$ in $\SN$. 
Formally, $\widehat{\N}^{k'}_{i}=\{v_j\mid d_\SN(v_j, v_i)=k'\}$. 
Note the difference of $\widehat{\N}^{k'}_{i}$ from $\N^{k'}_{i}$. 
As $\SN$ is connected, when $k'>0$, we have $\widehat{\N}^{k'}_{i}\subset \N^{k'}_{i}$.
The initial $H_j^0$ for each $v_j\in\V$ is set to $z^{\hist}_{j,t}$.

Note that different from~\cite{ZYZHAK20} which adopts the ${\sf Aggregate}$ 
function of GCN~\cite{KipfW17}, we use GAT~\cite{abs-1710-10903} for better performance. 
The output representation of $v_i$, denoted by $z^{\it social}_{i,t}$, is calculated 
by combining node representations of all layers:  
$$
z^{\it social}_{i,t} = {\sf Combine}(\{H^{0}_{i}, H^{1}_{i}, \dots, H^{k}_{i}\}).
$$ 
Many ways exist to implement the function ${\sf Combine}$. 
We adopt the one in H2GCN~\cite{ZYZHAK20} in our model which outputs
the concatenation of all inputs.

By concatenating $z^{\it social}_{i,t}$ and $z^{\Text}_{i,t}$, we obtain 
the text-level representation vector for classification. Formally,
$$ z_{i,t} = z^{\it social}_{i,t} \; \Vert \; z^{\Text}_{i,t}.$$

\smallskip\noindent
\textbf{Attitude classification.}
We implement a simple two-layer structure to  conduct the classification. 
Note the output vector has a length of $|\VA|$. Recall $\VA$ is the set of class labels.
The first layer applies the {\sf ReLU} function to each element in $z_{i,t}$ and 
at the second layer, we use a linear regression where 
$W\in\mathbb{R}^{2d\times |\VA|}$ and $b\in \mathbb{R}^{|\VA|}$ is the bias vector. 
The {\sf softmax} function is applied to calculate the probability distribution over the 
class labels. 
Formally, the distribution $p$ is calculated as follows:
$$p= {\sf softmax}\left({\sf ReLU}(z_{i,t})\cdot W + b \right).$$
The class label with the largest probability will be chosen as the output.
We use CrossEntropyLoss as the objective function to train the entire model.

\section{Data Curation}
\label{sec:data}
To train and evaluate our vaccination attitude learning framework, 
we collect a dataset from Twitter focusing on four adjacent Western European 
countries: Germany, France, Luxembourg and Belgium. 
Reasons to select these countries include their importance to 
European economy and their similar and almost synchronous pandemic management policies.
They can also well represent the first group of 
countries receiving and administering  COVID-19 vaccines. 
%
Due to the lack of publicly available data with vaccination stances, 
we create the first public set of annotated tweets for training and testing our framework. 
The statistics about our dataset are summarised in Table~\ref{tab:dataset statistics}.
This dataset and our annotation are publicly available.\footnote{The download link is hidden 
due to the double-blind review policy.}  

\begin{table}[!t]
\centering
\caption{Dataset statistics.}
\vspace{-2mm}
\label{tab:dataset statistics}
\resizebox{0.85\linewidth}{!}{
\begin{tabular}{|l|l|l|r|}
\hline
\multirow{8}{*}{\textbf{Social network}} & \multirow{5}{*}{\textbf{\#node}} &  France & 35,081 \\ 
\cline{3-4} 
 & & {Germany} & 16,304 \\ 
 \cline{3-4} 
 & & {Belgium} & 15,647 \\ 
 \cline{3-4} 
 & & {Luxembourg} & 2,904 \\ 
 \cline{3-4} 
 & &\multicolumn{2}{r|} {\textbf{69,936}} \\ 
 \cline{2-4} 
 & \textbf{\#edge} &\multicolumn{2}{r|}{8,909,985} \\ 
 \cline{2-4} 
 & \textbf{avg. degree} & \multicolumn{2}{r|}{127.40} \\ 
 \hline\hline
\multirow{6}{*}{\textbf{Tweets}} & \multirow{5}{*}{\textbf{\#tweet}} &  France & 5,925,354 \\ 
\cline{3-4} 
 & & {Germany} & 2,669,875 \\ 
 \cline{3-4} 
 & & {Belgium} & 530,885 \\ 
 \cline{3-4} 
 & & {Luxembourg} & 9,279 \\ 
 \cline{3-4} 
 & &\multicolumn{2}{r|} {\textbf{9,135,393}} \\ 
 \cline{2-4} 
 & \textbf{\#tweet/user} & \multicolumn{2}{r|}{130.62} \\ 
 \hline
 \hline
{\multirow{2}{1cm}{\textbf{Annotated tweets}}} & \textbf{\#tweet} & \multicolumn{2}{r|}{ 18,246}\\ 
\cline{2-4} 
 & \textbf{\#user} &\multicolumn{2}{r|}{4,157}  \\ 
 \hline
\end{tabular}
}
\vspace{-4mm}
\end{table}

\subsection{Data collection and preprocessing}
Our dataset consists of two types of data: i) a social network composed of active Twitter users, and ii)  
the tweets of selected users related to COVID-19 vaccine or vaccination. 
By `active users', we mean users that are active in 
vaccination-related discussion and frequently interact with others. 

\smallskip\noindent
\textbf{Step 1. Social graph construction.}
We start with identifying the Twitter users in our targeted region who actively participated 
in vaccine-related discourse. 
Instead of directly searching tweets by keywords, we refer to a publicly available dataset 
which contains the IDs of COVID-19 related tweets~\cite{COVID-19Dataset}. 
We extract the tweet IDs spanning between  
January 22, 2020 and March 15, 2021 covering the beginning of 
the vaccination campaign.  
Through these IDs, we download the corresponding tweets. 
Each downloaded tweet is associated with meta-information which 
includes the location of the originator, either self-provided 
by the originator or attached by the device's positioning services such as GPS.
Due to the ambiguity of the self-reported locations, 
we use the geocoding APIs, Geopy 
and  ArcGis Geocoding 
to regularise their formats. 
For example, a user input location \emph{Moselle} 
is transformed to a preciser and machine-parsable location: 
\emph{Mosselle, Lorraine, France}.
Based on the regularised locations, we filter the downloaded tweets and 
remove those posted by users out of the region. 
In total, we obtain 990,448 tweets from 767,583 users. 

To find the users with frequent interaction with other Twitter users,
we construct a re-tweeting weighted graph. 
An edge is created between  
two users if one user retweeted a tweet from the other user or mentioned him/her.
The edge weight is assigned as the number of mentions or retweets between them. 
We remove all edges with weights smaller than two and calculate the largest weakly 
connected component of the graph which consists of 72,960 active users.
As retweeting or mentioning a user does not mean these two users have a following relationship, 
we crawl the remaining users' followers, with which a
graph is constructed with the remaining users and the relationships between them as edges. 
In the end, we take the largest weakly connected component 
of the resulted graph as the final social graph. 
This graph consists of 69,936 nodes and 8,909,985 edges. On average, each user 
has 127.4 followers. This indicates our selected users are sufficiently active on Twitter.

\smallskip\noindent
\textbf{Step 2. Vaccine-related tweet collection.}
In this step, we crawl the tweets originated or retweeted by the users 
in the social graph. 
Note that we are only interested in the tweets related to COVID-19 vaccination.
We use a list of keywords to filter out the irrelevant ones. 
The keywords should be general enough to cover tweets in German, French 
and English. With our observation and several trials, we select the 
keywords containing the following strings: `\emph{vax}', `\emph{vaccin}', `\emph{covidvic}', 
`\emph{impfstoff}', `\emph{vacin}',  `\emph{vacuna}' and  `\emph{impfung}'.
We use the Twitter Academic Research API to download tweets. 
Due to the limitations of the API, 
each request can only download a maximum of 500 tweets. 
To be efficient and ensure the coverage ratio, we create a download request for every user in 
each month. In total, we collect 1,626,472 tweets, and each user 
has about 130 tweets on average. 
We clean the downloaded tweets by removing mentions of other users with `@', 
quoted hyperlinks and `RT'.

\begin{table}[!t]
\caption{Annotation labels and examples.} 
\vspace{-2mm}
\label{tab:annotation examples}
\centering
\resizebox{1.\linewidth}{!}{
\begin{tabular}{|l|p{7.2cm}|}
\hline
\textbf{Label}            & \textbf{Examples (translated to English) }                                                                       \\ \hline
\multirow{2}{*}{positive} & vaccines: why there are no long-term side effects.                                \\ \cline{2-2} 
                          & We have a new weapon against the virus: the vaccine. Hold together, again.       \\ \hline
\multirow{3}{*}{negative} &
  this nurse gets covid-19 vaccine. then she talks to media how great it is. then passes out. watch! \\ \cline{2-2} 
                          & Corona mass vaccination aka disability manufacturing                                       \\ 
\hline
\multirow{2}{*}{neutral}  & How safe is the Covid 19 vaccine for people with diabetes?                                                     \\ \cline{2-2} 
                          & If there is a covid vaccine, what will you do?                                            \\ \hline
\multirow{2}{2.5cm}{positive but dissatisfied with government management} 
                          & 
                          It's bad enough for individuals to refuse \#COVID19 \#vaccines for themselves. But forcing a mass vax site to shut down, knowing it means vaccines may go to waste, is criminal. Call it pandemicide.\\ \hline
\end{tabular}
}
 \vspace{-4mm}
\end{table}

\subsection{Data annotation}
According to the best of our knowledge, no datasets of social media posts are 
publicly available with users' COVID-19 vaccination attitudes annotated. 
As a result, we select a subset of our downloaded tweets and manually 
attach them with attitude labels.  In this paper, we focus on users' 
affective stances towards COVID-19 vaccination which are \emph{positive}, 
\emph{negative} and \emph{neutral}. Table~\ref{tab:annotation examples} lists
examples of the attitude labels.
After a closer check of the tweets, we notice a relatively large number of 
tweets in which users express their 
dissatisfaction or complaints about governments' COVID-19 management 
policies but possess positive attitudes towards vaccination.
Take the last row in Table~\ref{tab:annotation examples} for an example. 
It contains a few negative words such as `bad', `criminal' and `waste' but 
the originator explicitly expresses his/her support for vaccination.
As such tweets deliver a negative emotion, if we do not separate them 
from those with negative vaccination attitudes, NLP models will be confused 
and their classification accuracy will be deteriorated. 
Therefore, we add a label ${\it PD}$ indicating `\emph{positive but dissatisfied with 
government management}'. Then the set of labels $\VA$, used in the rest of 
the paper, consists of four labels, i.e., $\{{\it PO}, {\it NG}, {\it NE}, {\it PD}\}$.

We select tweets to be annotated with the purpose to cover all the users in 
our dataset. We order our downloaded
tweets in a list according to their numbers of times being re-tweeted in 
descending order. 
We then iteratively remove the most transmitted tweet from the list and 
put it into the list of selected tweets until every user 
in our dataset originally posted or retweeted at least one selected message. 
After this step, we select 18,246 tweets originated from 4,157 users.

\begin{figure}[t]
\centering
\includegraphics[width=0.32\textwidth]{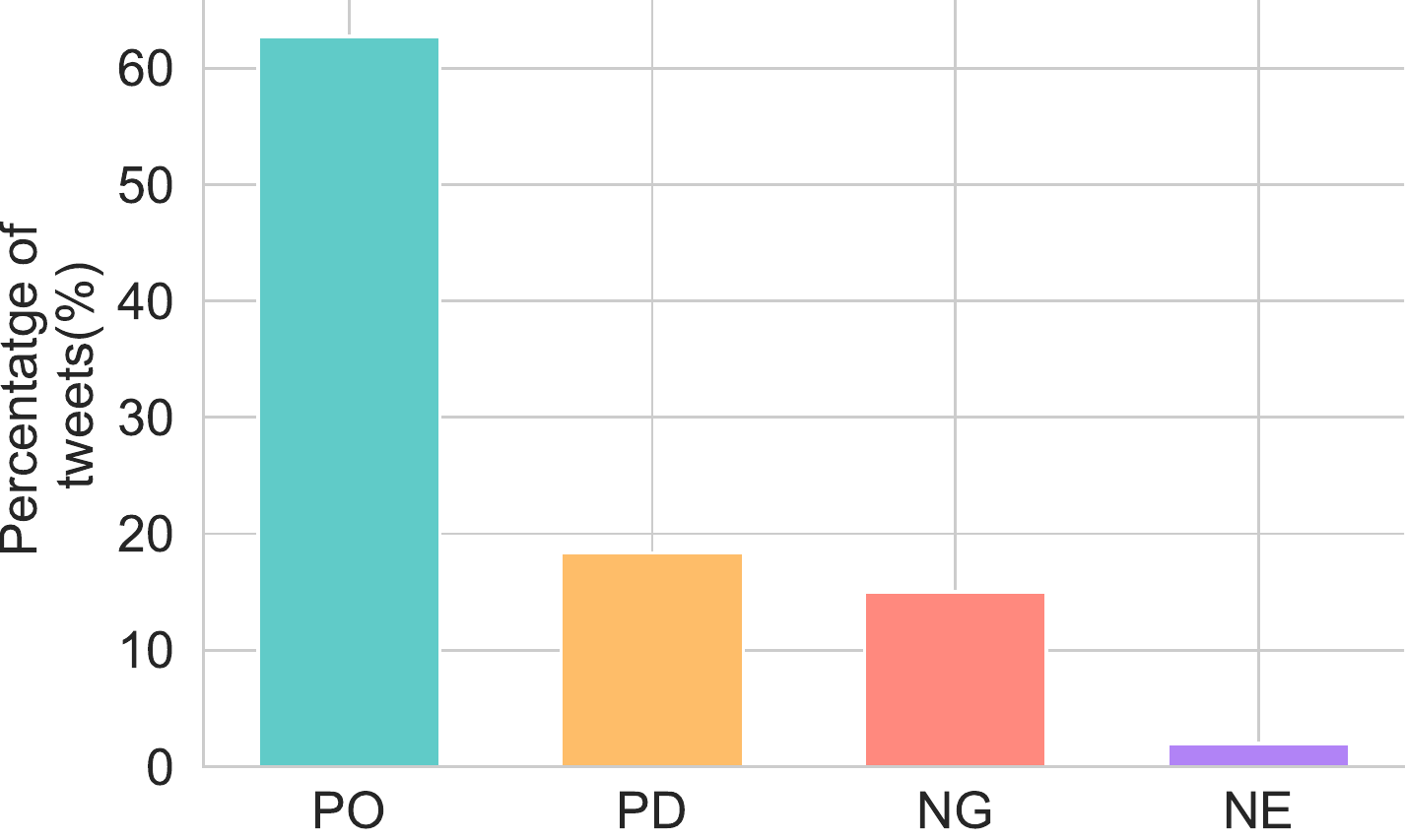}
\vspace{-2mm}
\caption{Distribution of vaccination attitude labels.}
\label{fig:label-distribution}
\vspace{-4mm}
\end{figure}

As our tweets are multilingual, we hire 10 bachelor students who can speak 
at least two of the three most used languages (i.e., German, French and English).
One author of this paper acts as the coordinator 
and trains all the hired annotators by explaining the 
semantics of all labels with examples. To ensure all annotators hold the same 
understanding of the labels, they are asked to annotate 200 tweets. 
The coordinator checks their annotation and communicates to them with extra
explanation when necessary. 
We conduct three rounds to make sure each tweet's label is double-validated. 
In the first round, an annotator annotates all the selected tweets. 
In the second round, each of the rest 9 annotators is assigned randomly 2,000 
tweets and validates the annotation. In the last round,  
the coordinator goes through all validated annotations. 
In Figure~\ref{fig:label-distribution}, we show the distributions of the tweets 
according to their labels. 

\paratitle{Annotator agreement.} We leverage three widely accepted 
measurements to evaluate the inter-annotator reliability for each label:
Average Observed Agreement (AOA)~\cite{fleiss2013statistical}, 
Fleiss' kappa~\cite{fleiss2013statistical}, and 
Krippendorff's Alpha~\cite{krippendorff1970estimating}. 
The values of all the three measurements range from 0 to 1, 
where 0 indicates complete disagreement and 1 indicates absolute agreement.
Table~\ref{tab:agreement} summarises the inter-annotator agreement for each annotation label. 
We can see that for labels {\it PO}, {\it NG} and {\it NE},  all the three measurements 
produce scores larger than 0.73, 
indicating an outstanding agreement level. 
The annotators' agreement on {\it PD} falls drastically compared to other labels, but still 
remains moderate according to the ranking criteria of the Fleiss' Kappa measurement. 
This can be explained by our difficulties during annotation in dealing with
the special linguist features of {\it PD} tweets, i.e., frequently used negative terms or 
sarcastic expression. 
\begin{table}[!t]
\caption{Inter-annotator agreement (\textbf{\textit{PO}}: Positive,
\textbf{\textit{NG}}: Negative, \textbf{\textit{NE}}: Neutral, 
\textbf{\textit{PD}}: Positive but dissatisfaction).} 
\label{tab:agreement}
\centering
\resizebox{0.9\linewidth}{!}{
{
\begin{tabular}{|l|r|r|r|}
\hline
\textbf{Label} & \multicolumn{1}{l|}{\textbf{AOA}} & \multicolumn{1}{l|}{\textbf{Feliss' kappa}} & \multicolumn{1}{l|}{\textbf{Krippen-dorf's Alpha}} \\ \hline\hline
{\it PO}                     & 0.72 & 0.73 & 0.73 \\ \hline
{\it NG}                    & 0.82 & 0.88 & 0.88 \\ \hline
{\it NE}                    & 0.74 & 0.78 & 0.77 \\ \hline
{\it PD} & 0.61 & 0.63 & 0.62 \\ \hline
\end{tabular}
}
}
\vspace{-4mm}
\end{table}

\section{Experimental evaluation}
\label{sec:experiments}
\paratitle{Evaluation setup.}
We set up an evaluation pipeline following the approach for traditional supervised 
classification~\cite{KM21}. 
Specifically, we split labelled tweets into training (80\%), validation (10\%) 
and testing (10\%) sets. 
The models are optimised with the training set, and the validation set is used to 
tune hyper-parameters. The model performance is evaluated on the testing set. 

\begin{figure}[!h]
\centering
\includegraphics[scale=0.28]{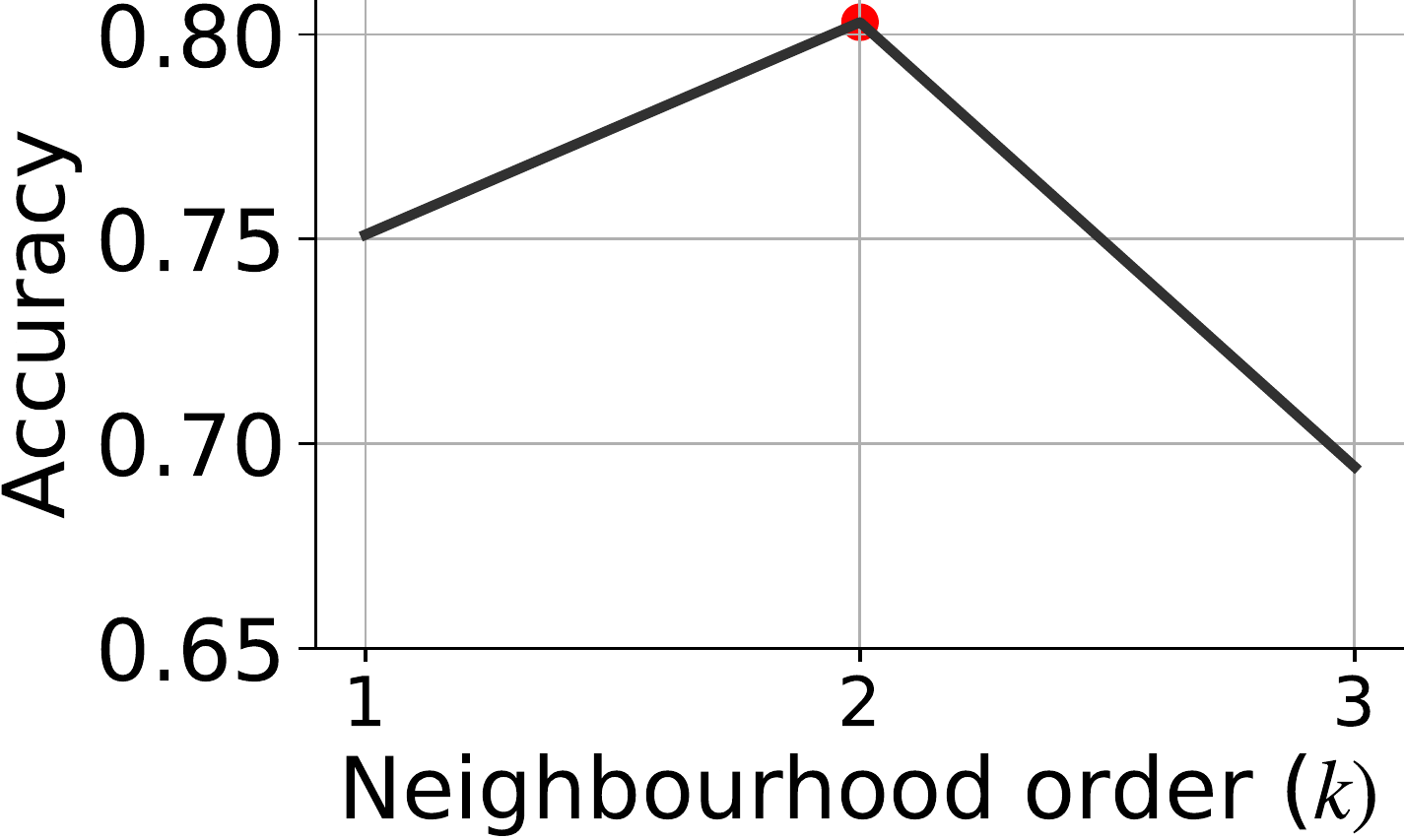}\quad
\includegraphics[scale=0.28]{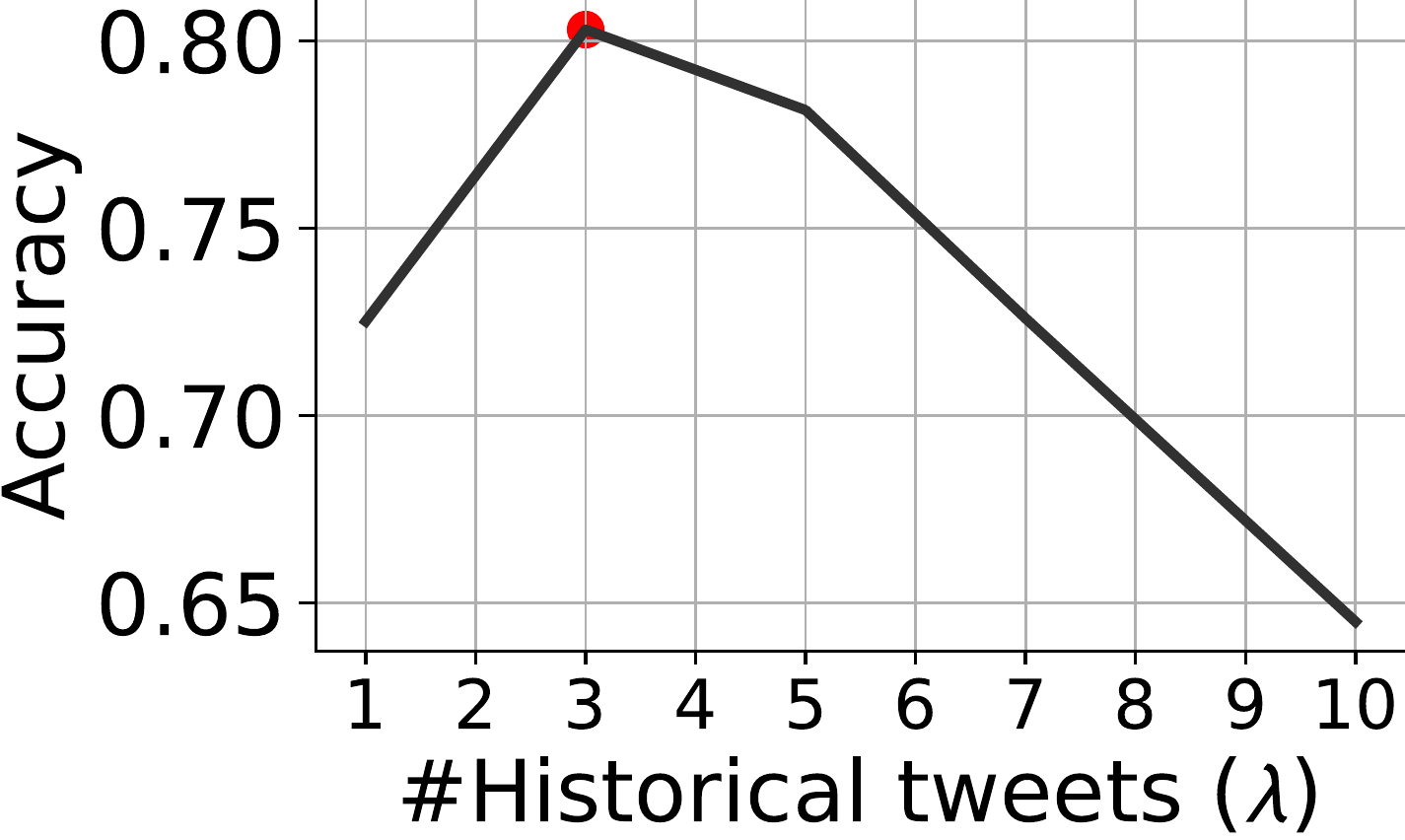}
\caption{Parameter tuning for $k$ and $\lambda$.}
\label{fig:parameter}
\vspace{-4mm}
\end{figure}

\paratitle{Hyperparameter settings.}
We train our model for 400 epochs and use Adam~\cite{KingmaB14} for optimisation with the 
learning rate of \num{e-5} and weight decay of \num{5e-4}.
For the text encoder, i.e., RoBERTa, we adopt the implementation XLM-RoBERTa~\cite{OL20} 
and follow their default settings where the maximum string length, i.e., parameter $d$, is 128. 
For our GNN-enhanced module,  we set the embedding dimension as $64$.
The neighbourhood order $k$ which is also the number of layers and 
the number of historical tweets $\lambda$ are important to ensure representation quality. 
Therefore, we conduct an empirical study to analyse the influence of these two key 
hyper-parameters to ensure the final performance. 
In Figure~\ref{fig:parameter}, we present the classification accuracy with different 
values of $k$ (on the left) and $\lambda$ (on the right).
We observe that these two hyper-parameters indeed significantly influence classification accuracy. 
Our model arrives at the best performance with $k=2$ and $\lambda=3$. 

\paratitle{Experimental results.} We compare our model with other possible 
implementations of our proposed vaccination attitude learning framework. 
In order to distinguish these models, we name them with two parts concatenated with `+'. 
The first part tells the adopted GNN model while the second part gives the method 
handling the temporal importance of past tweets. 
As all models use RoBERTa for text encoding, we do not explicitly put it in the model names. 
We present their performances in Table~\ref{tab:framework_results}. 

We have four major observations that justify the effectiveness of our implementation. 
First, the consideration of friends' vaccination discourse increases the performance. 
The text-only classification model with RoBERTa only has an average accuracy of 0.65 
while the other models, which are implemented with the GNN-enhanced module, achieve 
at least an accuracy above $0.70$. 
Second, the vaccination discourse between friends on Twitter is actually heterophily 
and the choice of heterophily-aware GNN models, i.e., H2GCN and our H2GAT, 
can further significantly improve the performance. 
The next four models below RoBERTa in Table~\ref{tab:framework_results} have the 
same settings except for the GNN methods. Both the application of H2GCN and our 
H2GAT achieve an increase of about 0.04 compared to the models with 
GCN~\cite{KipfW17} and GAT~\cite{abs-1710-10903}. 
Third, the consideration of the temporal importance of past tweets leads to 
another up to 0.06 improvement. We consider four methods to combine a 
user's last $\lambda$ tweets: MEAN, GRU, Hawkes and PE (short for positioning encoding). 
The method denoted by MEAN simply averages the text-level encodings. 
The positioning encoding method adopted in our model generates the best performance. 
Last, our extended H2GAT model outperforms the original H2GCN. 
Our implementation, i.e., H2GAT+PE, finally improves the text-only RoBERTa model 
by 23\% in terms of accuracy. 
\begin{table}[!t]
\caption{Model performance.\label{tab:framework_results}}
\vspace{-2mm}
\centering
\resizebox{1.\linewidth}{!}{
\begin{tabular}{|l|r|r|r|r|}
\hline
{\bf Model} & \multicolumn{1}{l|}{\bf Precision} & \multicolumn{1}{l|}{\bf Recall} & \multicolumn{1}{l|}{\bf F1} & \multicolumn{1}{l|}{\bf Accuracy} \\ \hline\hline
RoBERTa                  & $0.6758\ppm 0.0218$ & $0.5848\ppm 0.0232$ & $0.6249\ppm 0.0210$ & $0.6517 \ppm 0.0348$ \\ 
GCN+MEAN& $0.6936\ppm 0.0101$ & $0.6932\ppm 0.0153$ & $0.6890\ppm 0.0113$ & $0.7033\ppm 0.0087$  \\ 
GAT+MEAN& $0.7001\ppm 0.0091$ & $0.7002\ppm 0.0102$ & $0.6983\ppm 0.0078$ & $0.7096\ppm 0.0099$  \\ 
H2GCN+MEAN& $0.7387\ppm 0.0021$ & $0.7144\ppm 0.0081$ & $0.7286\ppm 0.0027$ & $0.7412\ppm 0.0015$  \\ 
H2GAT+MEAN& $0.7361\ppm 0.0038$ & $0.7371\ppm 0.0007$ & $0.7331\ppm 0.0019$ & $0.7461\ppm 0.0008$  \\ 
H2GCN+GRU    & $0.7813\ppm 0.0025$ & $0.7794\ppm 0.0048$ & $0.7712\ppm 0.0027$ & $0.7829\ppm 0.0010$  \\ 
H2GAT+GRU    & $0.7937\ppm 0.0036$ & $0.7968\ppm 0.0021$ & $0.7927\ppm 0.0011$ & $0.8009\ppm 0.0008$  \\ 
H2GCN+Hawkes & $0.7843\ppm 0.0016$ & $0.7760\ppm 0.0038$ & $0.7699\ppm 0.0047$ & $0.7831\ppm 0.0025$  \\ 
H2GAT+Hawkes & $0.7946\ppm 0.0036$ & $0.7922\ppm 0.0010$ & $0.7903\ppm 0.0019$ & $0.7988\ppm 0.0023$  \\ 
H2GCN+PE     & $0.7859\ppm 0.0031$ & $0.7813\ppm 0.0045$ & $0.7792\ppm 0.0022$ & $0.7889\ppm 0.0028$  \\ 
H2GAT+PE &
  $\textbf{0.7948}\ppm 0.0051$ &
  $\textbf{0.7954}\ppm 0.0029$ &
  $\textbf{0.7936}\ppm 0.0020$ &
  $\textbf{0.8030}\ppm 0.0017$ \\ \hline
\end{tabular}}
\vspace{-4mm}
\end{table}

\paratitle{Empirical complexity analysis.}
As the RoBERTa model is pre-trained, the models instantiated from our framework 
have the same complexity as the adopted GNN models. 
In our experiments, we conduct the training on a server with Xeon E5 CPU and Tesla V100 GPU.
On average, the training time for RoBERTA is about $115$ seconds for an epoch while 
$52.5$ seconds are needed for an epoch in training the GNN-enhanced module and the 
classification module.
What is more important is the running time of the models when processing a tweet. 
This will determine the practical utility of our framework in tracking 
public vaccination attitude in real time.  
We run four parallel instances of our model H2GAT+PE on the server.
On average, it takes 24.68 seconds for every 1,000 tweets, which means 
more than 3.5 million can be processed a day. 
For the regions we target at, we collect in total 9 million vaccination-related
tweets over two years. 
This implies our model is sufficiently efficient for processing posts 
on a daily basis.  

\paratitle{Cross-validation.} 
In addition to experimental evaluation, we also make use of published social 
studies to cross-validate our model's effectiveness. 
Lazarus et al. conducted a survey in June 2020, and estimated that the vaccine 
acceptance rates in France and Germany are 58.9\% and 64.5\%, respectively~\cite{lazarus2021global}. 
After applying our model to classify the tweets in the same period, 
we find the percentages of tweets with positive vaccination attitudes of 
these two countries are 42.27\% and 53.12\%, which are similar and 
retain the relative difference between the two countries.
This implies that posts on Twitter can be used as a reference 
to fast grasp the vaccine hesitancy levels when surveys are not available.

\begin{figure}[t]
\centering
 \includegraphics[width=0.9\linewidth]{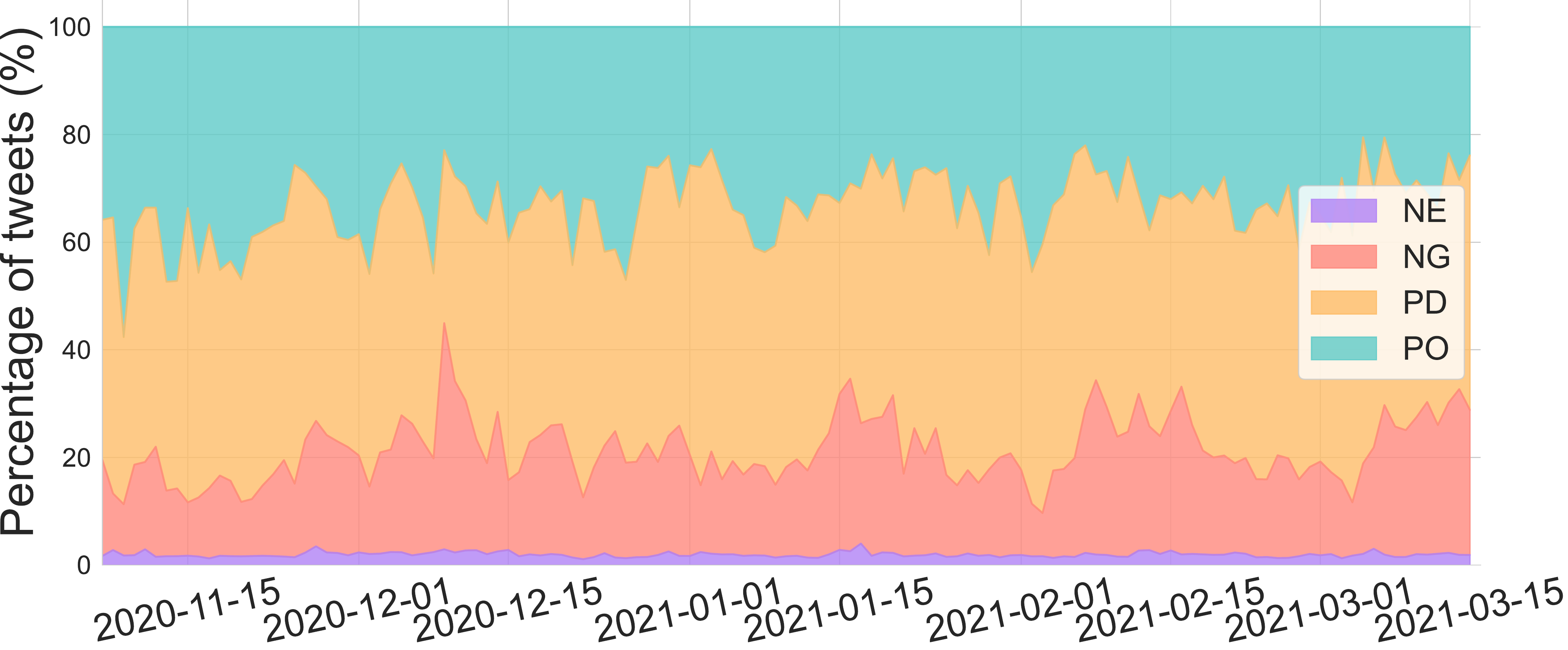}
\vspace{-2mm}
 \caption{Temporal distribution of tweets with different vaccination attitude labels.}
 \label{fig:temporal}
\vspace{-4mm}
\end{figure}

\begin{figure}[t]
\centering
\subfigure[Apex 1 (01/14--01/21)]
    {\includegraphics[width=0.32\linewidth]{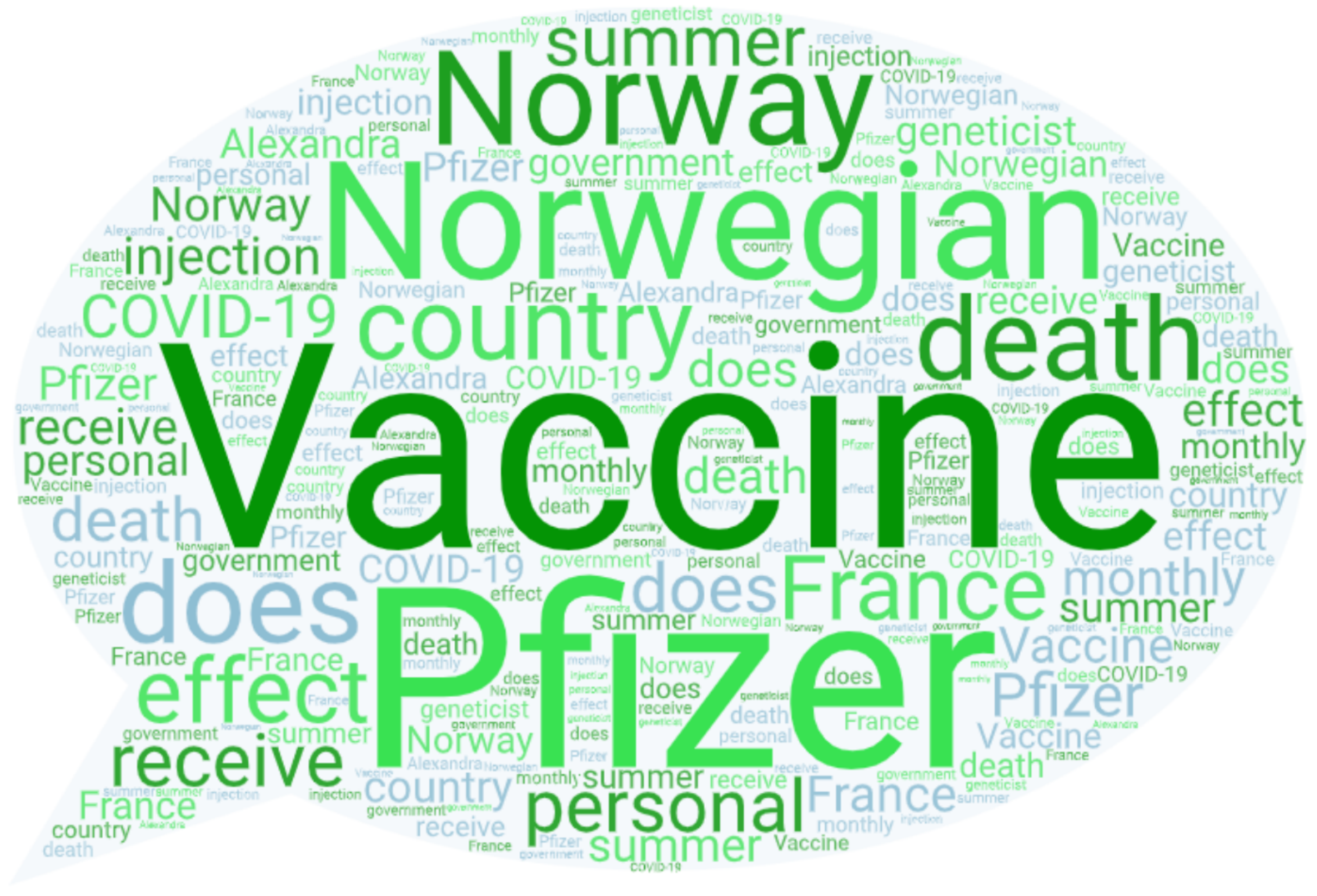}}
\subfigure[Apex 2 (02/10--02/17)]
    {\includegraphics[width=0.32\linewidth]{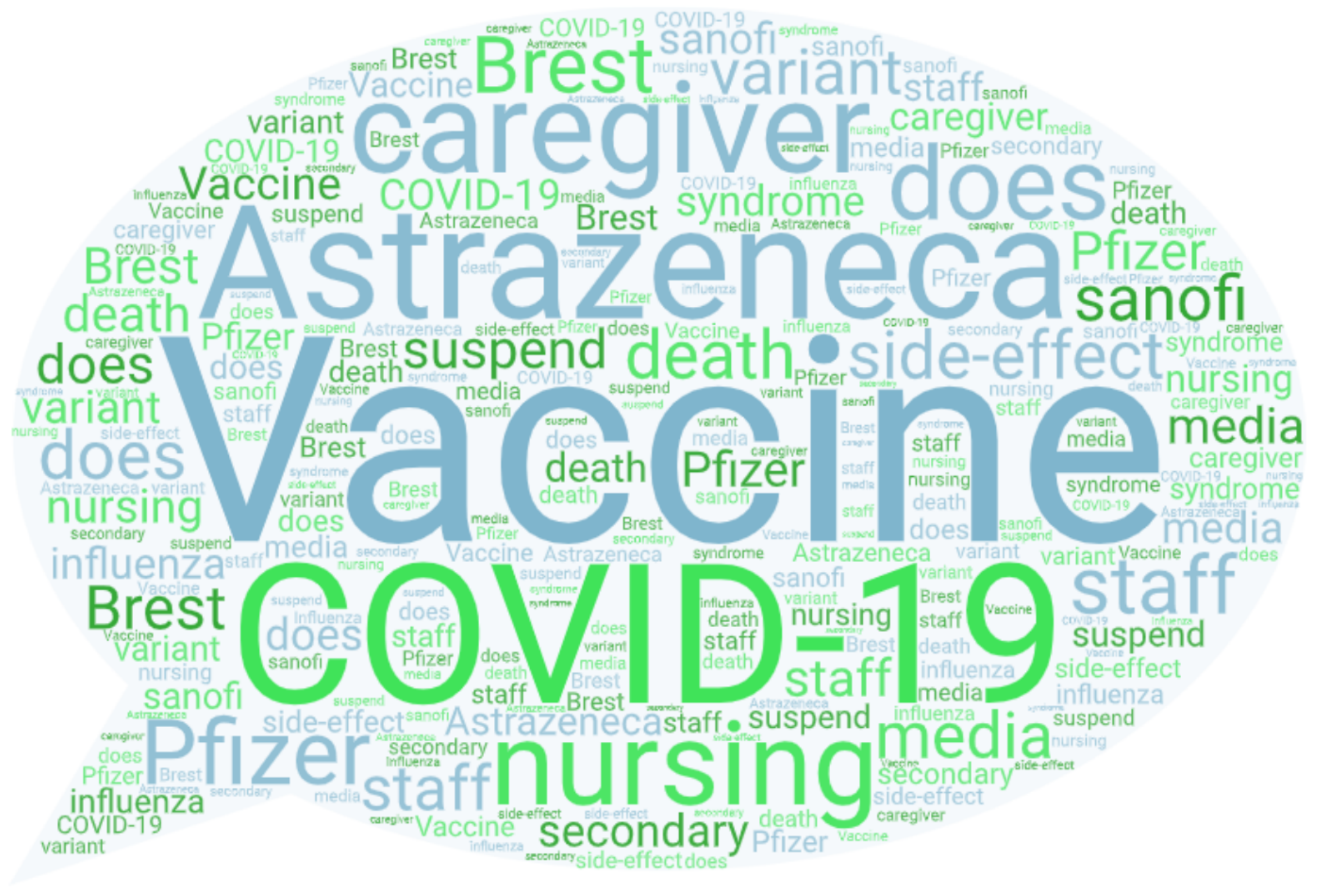}}
\subfigure[The base (01/30--02/06)\label{subfig:base}]
    {\includegraphics[width=0.32\linewidth]{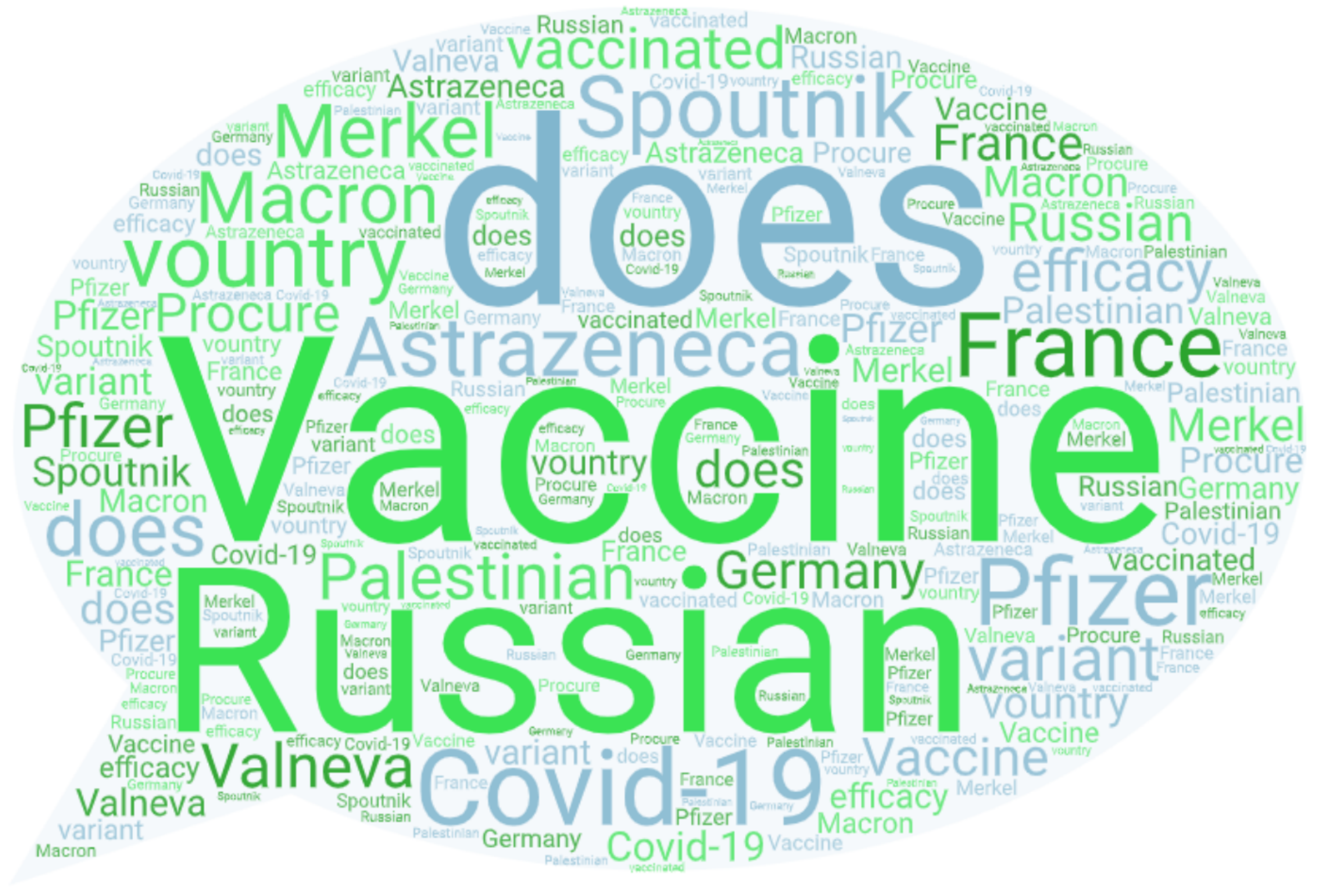}}
\vspace{-4mm}
\caption{Word clouds of tweets around selected points.}
\label{fig:word_cloud}
\vspace{-4mm}
\end{figure}

\paratitle{Vaccine hesitancy tracking and manual analysis.}
We draw the temporal evolution of the percentage of tweets for each selected label 
in Figure~\ref{fig:temporal} on a daily basis starting from November 8, 2020. 
Based on previous research reporting that the content of tweets 
is highly correlated with real-world situations~\cite{PCHG20}, 
we make a hypothesis that real-world events may contribute to 
the fluctuating proportion of tweets with different vaccination stances. 
In vaccine hesitancy monitoring, special attention should be paid to the fluctuations of 
negative attitudes. We take three time points that correspond to turning points of 
the curve of label {\it NE} as examples and discuss the potential causes. 
Among them, two correspond to apex points  where negative tweets 
reach local maximum percentages and one corresponds to a base point 
with local minimum negative tweets.
We first plot word clouds in Figure~\ref{fig:word_cloud} 
to identify the most frequently used keywords in the week around the selected points. 
Then we search these keywords on the Internet to identify the events that may contribute to the changes.

%
The first apex occurred around January 16, 2021.
We notice that this surge of negative tweets attributes to the propagation of 
a large volume of misinformation.
Take the two most dominant pieces of misinformation as examples. 
One said that on January 14, the Norwegian Medicines Agency reported that a total of 
29 people had suffered side effects, 13 of which were fatal.
The other was about the death of an Indian healthcare worker after receiving COVID-19 vaccines. 
The second peak happened around February 15, 2021.  
One piece of negative news was reported that AstraZeneca vaccines were stopped from administration 
after many health workers of Morlaix hospital in France 
suffered from side effects. This news subsequently led to anti-vaccination discussions.
The base point occurred between the two peaks around February 3, 2021. 
From Figure~\ref{subfig:base}, we find the dominant positive news that 
Russia started to offer other countries such as Pakistan 
with its vaccines. 

From the above discussion, we can see 
our model can enable the use of social media data to 
track on a daily basis the changes of vaccination attitudes, 
and capture the impact of social events on public vaccine hesitancy.
This may finally help the governments identify the right time 
to take intervention actions.

\section{Use case: Predicting vaccination hesitancy changes}
\label{sec:impact_analysis}
In this section, we illustrate a use of our vaccination attitude learning 
framework. Specifically, we analyse the role of the vaccination information widely spread 
across Twitter in affecting users' attitudes towards vaccination. 
Considering the comprehensiveness of vaccination discourses, 
we classify the most popular vaccination-related tweets into \emph{themes} that may correlate 
with vaccine hesitancy. Based on users'  perceived information 
in these themes, we succeed in forecasting their vaccination attitude changes with classic 
machine learning models.

\subsection{Period selection and theme labelling}
The participation of vaccination discourses fluctuates over time along with the occurrence 
of social events related to COVID-19 vaccines.  
We select two time periods after the start of COVID-19 vaccination campaign, 
in which the volume of tweets experiences significant surges compared to 
adjacent periods.
The first period lasts for 25 days spanning from December 27, 2020 to January 20, 2021 
while the second period lasts for 15 days between January 25 and February 8, 2021. 
These two periods involve 161,611 original tweets posted in total among which 
25,449 are retweeted at least once. 
The total number of times of being retweeted adds up to 242,129.  
We encounter two challenges to continue our analysis of the impact made by 
diffused information: the comprehensiveness and large volume of propagated tweets. 
Due to the huge volume of tweets propagated over Twitter, it is not plausible 
to consider all of them. 
Previous studies show that tweets' influence follows the power-law distribution
and 80\% of the impacts come from 20\% of the most widely spread
tweets~\cite{ediger2010massive}. 
Inspired by this result, we leverage the top 25\% most widely propagated 
tweets in every period to approximately represent the themes expressed in 
the diffused information. 
In total, we select 501 original tweets that are retweeted 78,891 times 
from 72.16\% of the users. To deal with the comprehensiveness,
with a careful examination of the selected tweets, we categorise them into 
themes that are considered to be responsible for the changes of vaccination 
attitudes. We refer to previous studies~\cite{bianco2019parent,della2021characteristics}, especially 
the Parent Attitudes about Childhood Vaccines (PACV) 
survey~\cite{opel2011validity} and the WHO Vaccine Hesitancy Matrix~\cite{world2015report}, 
and identify 11 themes that are relevant and can cover the propagated tweets
(see Table~\ref{tab:topic examples} in Appendix for explanation and examples).
We ask two of the 10 hired annotators to  manually annotate 
the selected tweets with their corresponding themes. 
The Cohen’s Kappa coefficient $k=0.82$ implies a high rate of agreement between them.

\subsection{Predictability of vaccine hesitancy changes}
\paratitle{Handling retweets and quotations.}
In addition to original posts, retweets and quotations also take up a large proportion of a user's 
historical posts. For quotations, a user added some comments which may express opposite opinions
to that of the quoted one. Therefore, we only use the texts users upload as valid
posts encoding users' vaccination stances. 
Although retweets cannot fully represent a user's own opinion, 
the behaviour of retweeting itself indicates some sort of agreement with the ideas
expressed in the message retweeted~\cite{DLZM15}. Based on this idea, we take  
retweets into account when calculating an individual user's vaccine hesitancy.  
The same approach is also adopted in the vaccination attitude tracking discussed 
in the previous section.

\paratitle{Quantifying individual vaccine hesitancy.}
We measure the vaccine hesitancy of an individual user according to 
the tweets posted or retweeted by the user 
in a time interval. Formally, it is calculated as:  
$\frac{N_p(v)-N_n(v)}{N_p(v)+N_n(v)}$
where $N_p(v)$ denotes the number of posts with positive vaccination 
attitudes of user $v$ during the selected interval, and $N_n(v)$ is the corresponding number of tweets with negative attitudes. 
Considering our purpose being idea validation, we do not distinguish the various significance of original posts and retweets. 

For each selected period, we use the tweets posted 14 days before and after the period 
to evaluate individual users' hesitancy levels and see how they change. 
In order to ensure the reliability, 
we only consider the users who posted or retweeted at least 3 tweets. 
If a user's vaccine hesitancy experiences a change smaller than $0.05$, 
we consider the user's attitude \emph{unchanged}, otherwise, \emph{increased} or 
\emph{decreased} depending on the change direction.

\begin{table}[!t]
\centering
\caption{Model performances for attitude change prediction.}
\vspace{-2mm}
\label{table:topcis performances}
\resizebox{0.9\linewidth}{!}{
\small
\begin{tabular}{|l|r|r|r|r|}
\hline
\textbf{Model} &
  \multicolumn{1}{c|}{\textbf{Precision}} &
  \multicolumn{1}{c|}{\textbf{Recall}} &
  \multicolumn{1}{c|}{\textbf{F1}} &
  \multicolumn{1}{c|}{\textbf{Accuracy}} \\ 
  \hline
\textbf{SVM}           & 0.7374 & 0.7382          & 0.7345          & 0.7477          \\
\textbf{Naive Bayes}   & 0.6468          & 0.6559          & 0.6427          & 0.6658          \\
\textbf{Random Forest} & 0.6811          & 0.6838          & 0.6795          & 0.6958          \\
\textbf{XGBoost}       & 0.7232          & 0.7229          & 0.7198          & 0.7342          \\
\textbf{GBDT}          & \textbf{0.7533} & \textbf{0.7516} & \textbf{0.7498} & \textbf{0.7603} \\ \hline
\end{tabular}
}
\vspace{-4mm}
\end{table}

\paratitle{Modelling perceived information.}
A Twitter user perceives information from the tweets retweeted or originally posted by 
his/her direct friends. As our focus is the information widely diffused on Twitter, 
we use a vector $I_u=(c_1, c_2, \ldots, c_m)$ to approximately represent a user $u$'s perceived information
where $c_i$ is the number of popular tweets a user receives from followers in $i$-th theme.
As we have $11$ themes, $m=11$ in our analysis.

\paratitle{Model evaluation.}
We make use of various standard machine learning methods to predict the change of a user 
$u$'s vaccination attitudes with the input of $I_u$. 
The methods consist of SVM with rbf kernel ($C=1$, $\gamma=0.1$), 
Naive Bayes ($\alpha=1$), Random Forest ($100$ trees with maximum tree depth of $5$), 
XGBoost ($100$ trees with maximum tree depth of $4$) and GBDT ($100$ trees with 
maximum tree depth of $5$). Table~\ref{table:topcis performances} shows the performance of 
these methods. All numbers are averaged over 5 training sessions. 
We can see all the methods can achieve reasonably good prediction performance
and GBDT outperforms the rest models with an accuracy of $0.76$.
When we consider additional factors such as users' vaccine hesitancy levels before the periods, 
the accuracy can be improved to $0.86$.

\paratitle{Discussion.} These results show that we can make accurate predictions with users' 
perceived popular information. 
Since we have empirically illustrated the plausibility to use social media posts to 
track public vaccination attitudes, 
the results imply that the diffused information on social media like Twitter
could be used as indicators to forecast the changes of vaccine hesitancy levels. 
As repeated many times, although such predictions cannot achieve the same level of 
trust as social surveys, they provide decision-makers with a method to quickly 
understand and get prepared for the potential damage of certain misinformation or compare 
different vaccine hesitancy intervention strategies over social media. 

%
%
%
%

\section{Conclusion and Discussion}
In this paper, we proposed a deep learning framework to learn vaccination 
attitudes from social media textual posts. Although vaccination attitudes extracted 
from social media cannot be as accurate and reliable as conventional social surveys, 
our framework allows for continuously tracking the fast development of 
public vaccination attitudes and capturing the changes that deserve 
specific attention in time. 
By leveraging friends' vaccination discourse as contextual information, 
our model successfully reduces the interference of linguistic features such as sarcasm and irony. 
Our model instantiated from the framework improves the state-of-the-art text-only method 
by up to 23\% in terms of accuracy according to our manually annotated dataset. 
With cross-validation with published statistics and manually analysis, 
we further validated the effectiveness of the model to capture public vaccine hesitancy  
in real life. 
After identifying 11 themes from widely diffused information on Twitter,
with the help of our model, we validated the predictability of 
users' vaccine hesitancy changes 
by the information they perceived from social media. 
This showed a potential use of our model in practice.
Through this paper, we established again the power of social media 
data in supplementing public health surveillance, especially in combating infectious virus
like COVID-19.

\paratitle{Limitations and future work.} 
We have three main limitations to address in future. 
First, in our work we have primarily focused on Twitter 
which potentially induces bias in our data and analysis.
Thus, it is important to extend our work to
other social media platforms such as Facebook and Instagram, and cross-validate
our results.
Second, we only analysed users' affective vaccination stances 
(i.e., positive, negative and neutral),
which can only be used as an indicator of users' intention of getting vaccinated.
It will be interesting to look deeper into users' tweets for a longer time 
and identify underlying determinants that lead to vaccination acceptance. 
Third, we used only the top 25\%  most widely spread tweets as representatives to 
extract the themes of diffused information partly limited by manual annotation. 
Some information in certain themes may be missed. As an interesting future work, 
we will develop effective NLP models to learn different tweet themes automatically. 

\paratitle{Ethical considerations.}
This work is based completely on public data and does not contain private 
information of individuals. Our dataset is built in accordance with the FAIR 
data principles~\cite{wilkinson2016fair} and Twitter Developer Agreement and Policy and related policies. 
Our release of the dataset is also compliant with General Data Protection Regulation (GDPR).
To conclude, we have no ethical violation in the collection and interpretation of data in our study.

\bibliographystyle{ACM-Reference-Format}
\bibliography{sample-base}

\clearpage
\appendix

\begin{table*}[hbt!]
\caption{Diffused information themes and examples.} 
\label{tab:topic examples}
\centering
\resizebox{1.0\linewidth}{!}{
{
\begin{tabular}{|p{3.8cm}|p{5.5cm}|p{8cm}|}
\hline
\textbf{Theme}                                                                               & \textbf{Description}                          & \textbf{Example (Translated to English)  }                                                                                                                                                                     \\ \hline\hline
 Positive news        & Positive news about vaccines \& vaccination & Pfizer/BioNTech's vaccine would be effective against the new British variant of COVID19.                                    \\ \hline
 Negative news & Negative news about vaccines \& vaccination & Portugal :  She dies 2 days after the vaccine (at 41 years old). Her family asks for explanations \\ \hline
Distrust in government management & 
Doubt about the trustworthiness of governments or medical institutions, e.g., regarding the daily update of statistics   
& They have lied to us so much about masks, chloroquine, contagion in children, that it will be difficult to trust them the day they will tell us about a harmless vaccine.   \\ \hline
 Dissatisfaction with politics/policies  & 
 Unsatisfactory views of politics/policies, such as ineffective vaccination programs.  & I am opposed to mandatory vaccination because all of the world's health organizations say that it is not the right way for a vaccine to spread.                     \\ \hline
Perception of the pharmaceutical industry     &      Perception that pharmaceutical manufacturers pursue only economic interests rather than public health interests              & Pfizer's CEO sold 60 percent of his shares when the Covid vaccine was announced. When the CEO sells, it stinks \\\hline                           Conspiracy    &   Content that describes the event as the secret acts of a powerful, malevolent force.  & 18 months they've been on the vaccine ???? When did they know there would be a Covid 19 "pandemic" ????           \\ \hline
Beliefs, attitudes
about health and
prevention  &  Personal views on vaccines and the immune system, e.g. homeopathy, natural immunity, alternative therapies. &There is no point in a generalized vaccine for a disease whose mortality is close to 0.05\%.   \\ \hline
Positive personal expression &    Personal expression of positive attitude towards vaccines       &  We have a new weapon against the virus: the vaccine. Hold together, again.     \\ \hline

Negative personal expression & Personal expression of negative attitude towards vaccines& Why could actually 1.5 billion Chinese get healthy without vaccination, and with us it only works with vaccination...?      \\ \hline
Positive information       &   Positive expressions about vaccines from healthcare professionals                   & \#COVID19 \#vaccinationHow does an mRNA vaccine work?    \\ \hline
 Negative information          &     Negative expressions about vaccines from healthcare professionals         &  My daughter, a nurse at the AP-HP, on the vaccine "Ah ah ah! They don't even dream about it, they start with the old ones so that we can attribute the side effects to age". \\ \hline
\end{tabular}
}
}
\end{table*}

\end{document}